\documentclass[aps,pre,reprint,amsmath,amssymb]{revtex4-1}
\usepackage{graphicx}
\usepackage{color}
\usepackage{epsfig}
\usepackage{amsmath,amssymb}
\usepackage{mathtools}
\renewcommand{\tilde}{\widetilde}

\begin{document}

\title{Electrohydrodynamics within electrical double layer in a pressure-driven flow in presence of finite temperature gradients}

\author{Tanmay Ghonge$^1$, Jeevanjyoti Chakraborty$^{2}$, Ranabir Dey$^1$, Suman Chakraborty$^{1,2}$}
\affiliation{$^1$Mechanical Engineering Department, Indian Institute of Technology Kharagpur, Kharagpur - 721302, India \\ $^2$Advanced Technology Development Centre, Indian Institute of Technology Kharagpur, Kharagpur - 721302, India}

\begin{abstract}
A wide spectrum of electrokinetic studies is modelled as isothermal ones to expedite analysis even when such conditions may be extremely difficult to realize in practice. As a clear and novel departure from this trend, we address the case of flow-induced electrohydrodynamics, commonly referred to as streaming potential, in a situation where finite temperature gradients do indeed exist. By way of analysing a model problem of flow through a narrow parallel plate channel, we show that the temperature gradients have a significant effect on the streaming potential, and, consequently, on the flow itself. We incorporate thermoelectric effects in our model by a full-fledged coupling among the electric potential, the ionic species distribution, the fluid velocity and the local fluid temperature fields without resorting to ad hoc simplifications. We expect this expository study to contribute towards more sophisticated future inquiries into practical micro-/nano-fluidic applications coupling thermal field focusing with electrokinetic effects.
\end{abstract}

\date{\today}
\maketitle

\section{Introduction}

Streaming potential is one of the four primary electrokinetic phenomena, the other three being electroosmosis, electrophoresis and sedimentation potential. The genesis of these phenomena is contingent on the development of an electrical double layer (EDL) which refers to the space charge distribution in a dielectric medium together with the electrified surface in whose immediate vicinity such distribution is established through a balance between Coulombic and entropic interactions \cite{HunterBookNew}. The particular phenomenon of streaming potential is, however, set apart by the fact that its manifestation does not depend on the application of an external field (unlike electroosmosis and electrophoresis) nor does it involve the transport of particles bearing such electrified surfaces (unlike electrophoresis and sedimentation potential). As long as flow (actuated through simple mechanical actuation) of fluid, bearing the space charge distribution, takes place past the electrified surface, a streaming current will necessarily be manifested, and given a scope of charge accumulation through the specific geometry, a streaming potential may also be expected to be manifested together with such current. 

Indeed, it is because of this apparent simplicity that ever since the discovery of this phenomenon more than 150 years ago by Quincke \cite{1861AnnPhysQuincke}, streaming potential has been found to be a key element in the explanation of various phenomena in areas as diverse as physiological \cite{2009AnnRevFluidMechMechanotransduction, 2009JOrthopResSP, 2011AnnRevMaterResGrodzinsky} to geophysical \cite{2007AnnRevEarthPlanetSciSP}. It has also been used in a host of applications in the colloidal science realm; for instance, zeta potential measurement and electrokinetic characterization of surfaces \cite{1980JCISWagenenAndrade, 1998JCISDukhin, 2001JCISEricksonLi, 2010LangmuirIsraelachvili, 2012PCCPGallardoMoreno}. Most recently, it has been at the forefront in the world-wide research agenda of innovative energy conversion techniques \cite{1964JApplMechOsterle, 1965JChemPhysOsterle, 1975JHydroSrivastava, 2003JMicromechMicroengYang, 2004NanoLettMajumdar, 2005SensActBEijkel, 2005JNanoscienceNanotechYang, 2006NanoLettHeyden, 2006JMicromechMicroengKarnikMajumdar, 2007NanoLettHeyden, 2007LocArticle, 2008JPowerSourcesXuan, 2008EPXuan, 2008NanotechStein, 2008APLXie, 2008JPhysChemCDuffin, 2009JPowerSourcesChein, 2009POFFarooqueSumanSir, 2010MNWang, 2010MNChang, 2010JCISBerli, 2010EPChein, 2010LangmuirPrakashDaSumanSir, 2010EPGaraiSumanSir, 2011LangmuirAdityaSumanSir, 2012NanoLettGillespie, 2012APLAdityaSumanSir, 2012LOCSubirSir}. Nevertheless, despite the long and what would otherwise seem an `established' history \cite{1948RoyalElton1, 1949RoyalElton4, 1964JPhysChemBurgreenNakache, 1965JPhysChemRiceWhitehead, 1975JCISLevine, 1986JCISDonathVoigt}, ongoing research efforts continue to further our fundamental understanding as well to extend the possibilities of its use in tandem with newer surface and flow characteristics \cite{2007POFSherwood, 2008LangmuirSherwood, 2008PREChakrabortyDas, 2008MNXuan, 2009POFSherwood, 2009JFMSherwood, 2009JCISParkLim, 2009LangmuirSiddharthaDaSumanSir, 2009JChemPhysTamalDaSiddharthaDaSumanSir, 2010JChemPhysWang, 2010LangmuirSiddharthaDaSumanSir, 2010POFJeevanSumanSir, 2011JFMSchnitzerYarivFrankel, 2011POFJeevanSumanSir, 2011PRLSchnitzerYarivKhair, 2011POFZhao, 2011JACSBipolar, 2012EPJeevanSBRSirSumanSir, 2012JFMSchnitzerFrankelYariv, 2012PREJeevanRanabirSumanSir}. 

Interestingly, such research and modeling efforts have been carried out, almost without exception, with the unquestioned assumption that the systems under consideration are isothermal. This is despite the ubiquity of non-isothermal natural settings, and also inspite of the fact that even in laboratory settings, perfect isothermal conditions are difficult to realize in practice. Such an approach is paradigmatic of most of the electrokinetic modeling efforts inasmuch as it pertains to the micro- and nano-fluidic context. To be sure, there do exist numerous works which are indeed concerned with variations of temperature. However, these involve simplistic one-way couplings \cite{1997IJHMTDongqingLi, 2003JHTMaynesWebb, 2006IJHMTSumanSir, 2006NumHTTan, 2009IJHMTElazhary, 2010IJHMTSadeghi}, or, at best, couplings between momentum and energy through temperature dependent thermophysical properties \cite{2008IJHMTHwang, 2009JCISKwak, 2012IJHMTSadeghiSumanSir}. Importantly, they do not consider the fundamental dependence of the ionic fluxes on the temperature variation. This situation is rather surprising particularly when considered in the context of the rich theory that already exists to model coupled transport of momentum, heat and mass based on general non-equilibrium thermodynamic principles \cite{deGrootMazurBook, KondepudiPrigogineBook}. Such general theories have routinely been adapted to represent various transport phenomena involved in membrane technology \cite{1991JMembSciBaranowski}. Investigations involving non-isothermal transport are also quite common in electrochemical systems under the purview of thermoelectrochemistry \cite{NewmanBook}, particularly at high temperatures \cite{1997JElectroanalChemEngelhardt}. On another front, certain fundamental modelling frameworks have also been developed under the purview of colloidal science \cite{1987DerjaguinBook, 1989AnnRevFluidMechAnderson}. Despite the obvious commonalities that exist between these areas and micro and nano-fluidics, the necessity to model non-isothermal electrokinetic transport through these micro and nano-channels, in general, and streaming potential mediated non-isothermal transport in particular has remained largely unattended \cite{2012HeidelbergDietzelHardt}. 

It is only in the last few years that this situation has been remedied to a great extent through a series of works \cite{2005LangmuirPutnamCahill, 2008PRLWurger, 2008PRLGolestanian, 2008EPJEWurger, 2008PREWurger, 2009LangmuirWurger, 2009JPhysCondMattWurger, 2010RepProgPhysWurger, 2010LangmuirPiazza, 2011JChemPhysSeebeckcoeff, 2012PRLMajeeWurger} where the fundamental influences of temperature gradient on the flux of the ionic species itself have been explicitly captured following the strategy by Guthrie \textit{et al.} \cite{1949JChemPhysGuthrie} Notable precursors to these works were the ones by Ruckenstein \cite{1981JCISRuckenstein} and Morozov \cite{1999JETPMorozov}. These works have also motivated the application of such gradients for controlled manipulation of particle motion in microfluidic channels \cite{2010SoftMattStonePiazza}. While the modeling efforts in these works are directed towards the transport of colloidal particles and are, hence, reminiscent of the aforementioned earlier works \cite{1987DerjaguinBook, 1989AnnRevFluidMechAnderson}, important differences do exist in the approach itself \cite{2008PREWurger}. In contrast to the previous works which involve a dependence on the enthalpy in the velocity expression, these recent efforts express the velocity in terms of the temperature gradient in an explicit and straightforward way. This is done by taking into account the contribution of the Soret effect directly in the flux. Pertinently, the Soret effect refers to the motion of a particle when placed in a temperature gradient \cite{1856Ludwig, 1879Soret}. When the particle itself is charged (as it in the case of an ionic species) such transport may work in tandem with conventionally recognized responses to concentration and electrical potential gradients to generate novel flow characteristics. 

The influences of the rich interplay among these various factors under conditions of dynamic equilibrium are particularly intriguing in the case of streaming potential mediated flow. This is because it holds the possibility of unveiling unexpected trends in the otherwise routine, primarily pressure-gradient actuated flow through the simple application of a temperature gradient. This motivates the primary objective of the current study: To investigate the influence of an externally applied temperature gradient on a pressure-gradient driven flow which also results in the generation of a streaming potential. Notwithstanding the apparent simplicity of such an objective, the scope of this investigation is far-reaching. An insight into the complexities involved is in order.

The overarching condition for setting up the dynamic equilibrium is that the total ionic current across the cross-section of the flow conduit should necessarily be zero. In the absence of any temperature gradient, such a condition results in the manifestation of a streaming potential field through a balance of the streaming current and the conduction current. However, when a temperature gradient \emph{is} applied, the consideration of the Soret effect may alone add two levels of complexity. The first rather straightforward one is the motion of the particle in response to the temperature gradient: its direction depending on its thermophobicity or thermophilicity, and its intensity on the magnitude of the Soret diffusion coefficient which is effectively a function of the ionic heat of transport. The second level of complexity stems from the different values of such ionic heats of transport of the cations and the anions which leads to a separation of charge parallel to the direction in which the temperature is applied, and thus sets up its own potential difference. This is referred to as the Seebeck effect and the electric field associated with the potential difference is known as the thermoelectric field. The most crucial thing to note is that under the conditions of dynamic equilibrium, a certain potential difference \emph{is} generated along the flow direction. This potential difference is, however, devoid of any manifest identity of its genesis in either the Seebeck effect or the streaming potential phenomenon, and is solely a self-consistent manifestation of the combined interplay of the various mechanical, chemical, thermal and electrical factors. The fluid flow that is ultimately generated is thus a function of the applied pressure-gradient, the applied temperature gradient and the self-consistently generated electrical potential difference. 

Notwithstanding the deep synergy among the various factors at work, we are able to delineate the specific extent to which the Soret effect, in tandem with the Seebeck effect (that leads, in turn, to the thermoelectric field), influences the streaming potential mediated flow. The primary finding of this work is that with a sufficiently strong Seebeck effect, it may be possible to either augment or negate the volumetric suppression of the primary pressure-driven flow due to the streaming potential field. This is particularly important because such reversal of flow nature is achieved simply on the basis of the electrolyte nature without changing the applied temperature gradient.

The remaining part of this article is organized as follows. In Section II, we describe the the model problem as a way of analysing the influence of temperature gradient on streaming potential modulated flows. We also outline the general equations governing the electrical potential, the ionic species distribution (incorporating thermoelectric effects), the fluid-flow equations and the energy equation that need to be solved in a coupled way for a resolution of the intrinsic interdependence among the various fields. Additionally, in this section, we derive the electric field associated with the streaming potential incorporating such thermoelectric effects. In Section III, we adopt a non-dimensional scheme and present the dimensionless versions of the coupled equations of Section II. In Section IV, we discuss the boundary conditions. In Section V, we report and discuss the major findings of our investigation. Finally, in Section VI, we draw important conclusions based on these findings.


\section{Mathematical Formulation}
\begin{figure}[t]
\includegraphics[width=8cm]{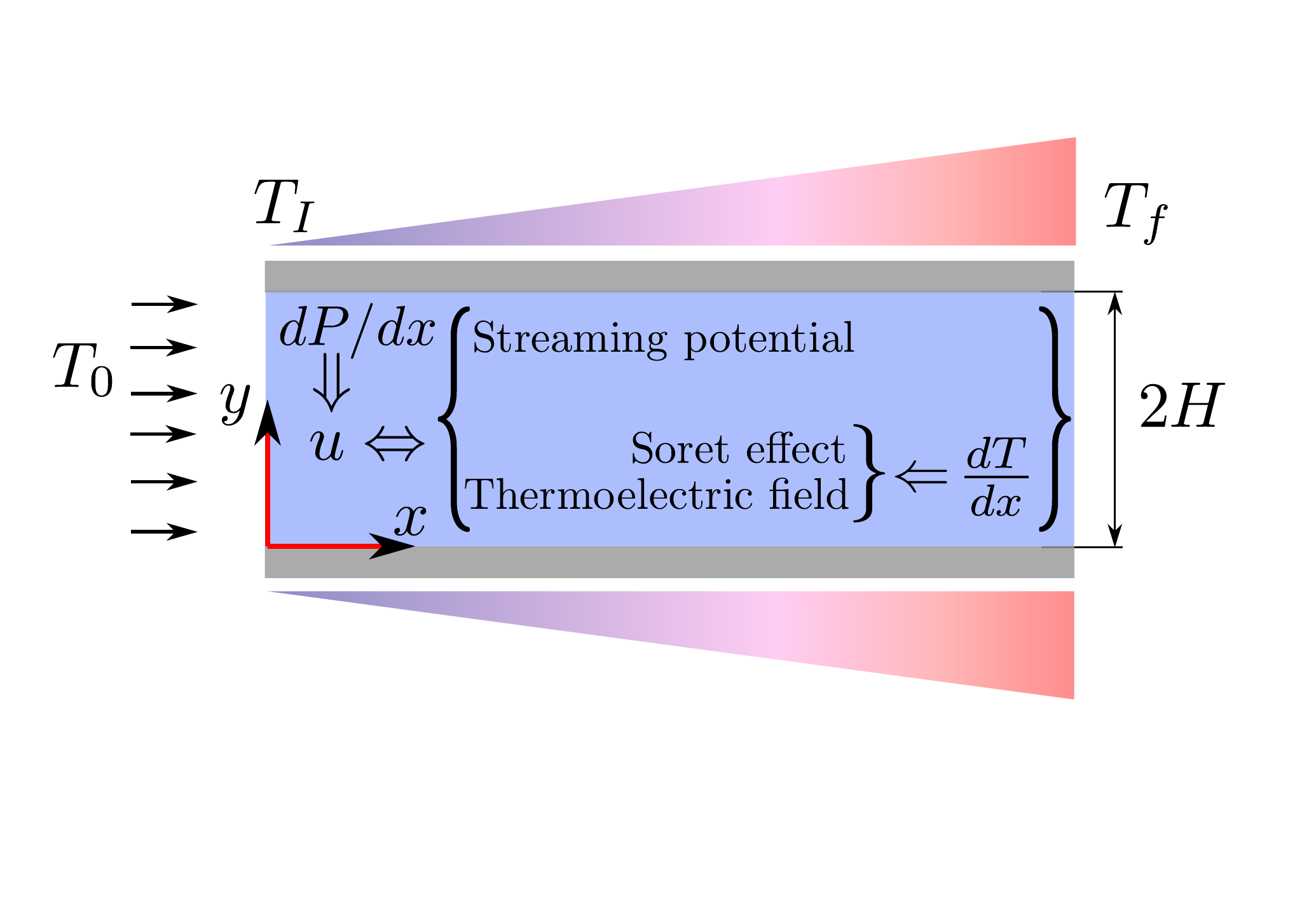}
\caption{Schematic of the problem geometry}
\end{figure}
Pressure-driven transport of a Newtonian fluid, containing symmetric electrolytes ($z_+ = z_- = -z$), through a long, parallel-plate channel of height $2H$, length $L$ and width $W$ ($W>>2H$) is considered here. The flow is actuated by a constant axial pressure-gradient, $P_x = -dP/dx$. The fluid enters the channel at temperature $T_0$ with a uniform velocity. The tip of the plate where the fluid enters is maintained at temperature $T_1$. There is an imposed linear temperature gradient on both the plates. The temperature of the tip of the channel where the fluid leaves is $T_f$. To  simplify the problem the following assumptions are made: steady, incompressible, and laminar flow of the electrolyte is considered, the thermophysical properties of the fluid are considered to be independent of temperature variations, the ionic species are assumed to behave as point charges, and the zeta potential, $\zeta$, is uniform along the channel walls.

\subsection{Potential distribution}
From the theory of electrostatics the potential (due to electrokinetic potential and thermoelectric potential), $\psi$, in the diffuse layer of EDL is governed by the Poisson equation which can be described as:
\begin{eqnarray}
\nabla^2\psi = -\frac{\rho_e}{\epsilon},
\end{eqnarray}
where $\epsilon$ is the permittivity of the medium. For $z:z$ symmetric electrolyte, the charge density, $\rho_e$ may be expressed as:
\begin{eqnarray}
\rho_e = ez \left( n_+ - n_- \right),
\end{eqnarray}
where $e$ is the electronic charge magnitude, $n_+$ and $n_-$ are, respectively, the number density of the positive ions and the negative ions. Then the Poisson equation may be written as:
\begin{eqnarray}
\nabla^2\psi = - \frac{ez \left( n_+ - n_-\right)}{\epsilon} \label{eq:Poisson}
\end{eqnarray}

\subsection{Species Transport}
The general species transport equation routinely invoked in continuum descriptions of electrokinetic phenomena is:
\begin{eqnarray}
\frac{\partial n_\pm}{\partial t} = - \nabla \cdot \mathbf{j}_\pm,
\end{eqnarray}
where $\mathbf{j}_\pm$ is the flux of the cation and the anion. Here, an important assumption is that there is no source term due to generation or consumption of any ionic species by any bulk reaction within the electrolyte. In conventional isothermal treatments, this flux arises as a combination of diffusion, electromigration and advection \cite{ProbsteinBook}. However, in the present case, there is an additional contribution to the flux from the temperature gradient. Physically, this comes about due to the propensity of a species to move in response to gradients in the temperature field, and is referred to as the Soret effect. This has rich implications in the resulting physical scenario. For, this motion of the ionic species due to the temperature field may itself result in the generation of a potential especially when the thermal diffusivities of the ionic species, with different polarities of charge on them, are different. The generation of a thermoelectric potential is analogous to the Seebeck effect. The total electric potential which influences the flux is, thus, a combination of the intrinsic electrokinetic screening potential and the thermoelectric potential together with the streaming potential which is induced parallel to the flow direction as a result of the streaming motion of the ions associated with the fluid flow . The resultant species transport is, therefore, an intimately coupled manifestation of this combined electric potential, the temperature gradient, the gradients in the concentration (which is itself coupled to the potential through the Poisson equation), and the advection association with the fluid motion (which, as we show later, is, again, influenced by the electric potential and concentration gradients). To address this highly coupled phenomenon, we first note that the flux of the ionic species may be expressed as \cite{2008PRLWurger}: 
\begin{eqnarray}
\mathbf{j}_+  = n_+ \mathbf{u}  - D_+\left[\nabla n_+ + \frac{n_+Q_+}{k_B T^2}\nabla T + \frac{n_+ez}{k_B T}\left( \nabla \psi  - \mathbf{E}_s \right)\right], \nonumber\\
{}\\ 
\mathbf{j}_-  = n_- \mathbf{u}  - D_-\left[\nabla n_- + \frac{n_-Q_-}{k_B T^2}\nabla T - \frac{n_-ez}{k_B T}\left( \nabla \psi  - \mathbf{E}_s \right)\right], \nonumber\\
\end{eqnarray}
where $\mathbf{u}$ is the mean fluid velocity, $k_B$ is the Boltzmann constant, $T$ is the absolute temperature, $\mathbf{E}_s$ is the electric field associated with the streaming potential, $D_\pm$ denote the diffusitivities of the posistive and the negative ions, and $Q_\pm$ denote the ionic heat of transport of the positive and the negative ions. Then, at steady state, the transport equations for ions may be written as:  
\begin{widetext}
\begin{eqnarray}
\nabla  \cdot \left[ D_+\left\{\nabla n_+ + \frac{n_+ Q_+}{k_B T^2}\nabla T + \frac{n_+ez}{k_B T}\left(\nabla \psi - \mathbf{E}_s \right) \right\} - n_+ \mathbf{u} \right] = 0,  \label{eq:nplus_tp} \\
\nabla  \cdot \left[ D_-\left\{\nabla n_- + \frac{n_- Q_-}{k_B T^2}\nabla T - \frac{n_-ez}{k_B T}\left(\nabla \psi - \mathbf{E}_s \right) \right\} - n_- \mathbf{u} \right] = 0.  \label{eq:nminus_tp}
\end{eqnarray}
\end{widetext}
\subsection{Streaming Potential Field}
The downstream migration of ions due to the flow primarily actuated by the imposed pressure gradient gives rise to a current known as the streaming current ($I_s$). However, in the stationary state, this convective transport of ions sets up its own electric potential known as the streaming potential. The electric field ($E_s$) associated with this streaming potential generates a current, known as the conduction current ($I_c$). Much like the electroosmotic flow situation, this conduction current, in turn, leads to a fluid flow opposite to the pressure-driven flow which was responsible for inducing the same in the first place. In the absence of an imposed electric field, the conduction current and the streaming current must balance each other so that the net ionic current in the system along the axial direction is zero; thus:
\begin{eqnarray}
I_{ionic} = I_s + I_c = 0.
\end{eqnarray}
It is important to realize that the real physical situation is far more involved that this rather simplistic description might apparently indicate. This is because of the inricate coupling among the electric potential (with contributions from the electrokinetic screening potential, the thermoelectric potential and the streaming potential), the concentration profiles, the temperature and the velocity fields. Such an intricate coupling might indeed include situations where locally the flow might be in a direction opposite to that expected from the primary pressure gradient drive. The only requirement that needs to be necessarily satisfied is that the total ionic current along the axial direction integrated over any cross-section of the channel vanishes. In order to capture this general requirement, we express the local ionic current in terms of the axial direction components of the cationic and the anionic fluxes:
%
%
%
\begin{eqnarray}
i=ez(j_{+x} - j_{-x}),
\end{eqnarray}
so that the total ionic current across a cross-section of the channel is given by:
\begin{eqnarray}
I_{ionic} = e z \int^{2H}_0 \left( j_{+x} - j_{-x} \right) dy. 
\end{eqnarray}
Thereafter, imposing the condition of the vanishing total ionic current, we obtain an expression for the electric field, $E_s$ associated with the streaming potential in the form:
\begin{widetext}
\begin{eqnarray}
E_s = \frac{\int_0^{2H}\left( n_+ - n_- \right) u dy - \int_0^{2H} \left[ D_+ \frac{\partial n_+}{\partial x} - D_- \frac{\partial n_-}{\partial x} \right]dy - \frac{ez}{k_B}\int_0^{2H} \frac{D_+ n_+ + D_- n_-}{T} \frac{\partial \psi}{\partial x}dy - \int_0^{2H} \left[ \frac{D_+ n_+ Q_+ - D_- n_- Q_-}{k_BT^2} \right]\frac{\partial T}{\partial x} dy}{-\frac{ez}{k_B}\int_0^{2H}\frac{D_+n_+ + D_-n_-}{T}dy}. \label{eq:Es} \nonumber \\
\end{eqnarray}
\end{widetext}
The important thing to note here is that this expression is not an explicit expression for $E_s$ because the velocity field in the first term in the numerator depends on this streaming potential field itself. To obtain a proper estimate of this dependence, we next move on to a description of the velocity fields from the Navier-Stokes equations representing the momentum transport of the fluid. 

\subsection{Navier-Stokes Equation}
The axial direction velocity contributing to the flux and, hence, to the streaming potential field is governed by the x-momentum equation which under the assumption of low Reynolds number flow reduces to:
\begin{eqnarray}
0 = \frac{ - dP}{dx} + \mu \frac{\partial ^2 u}{\partial y^2} + \mu \frac{\partial ^2 u}{\partial x^2} + F_x.
\end{eqnarray}
where the total body force $F_x$ is made up of two contributions: First, $F_{O_x}$ due to the osmotic pressure, and, second, $F_{E_x}$ due to the Maxwell stress along the axial direction. Noting that the general expression of force due to osmotic pressure is:
\begin{eqnarray}
\mathbf{F}_O = - \nabla \left( n k_B T \right)
\end{eqnarray}
with $n=n_+ + n_-$, the x-component of this force is:
\begin{eqnarray}
F_{O_x} =  - k_B(n_+ + n_-)\frac{\partial T}{\partial x} - k_B T\frac{\partial (n_+ + n_-)}{\partial x} \label{eq:FO_ax}
\end{eqnarray}
It is, therefore, this contribution from the osmotic pressure which takes into account the dependence of the velocity field on the imposed temperature gradient and the concentration gradients that are established as a combined consequence of the electrokinetic and thermoelectric effects. It is important to note that there does exist an alternative route to this explicit consideration of osmotic pressure contributions separate from purely hydrodynamic pressure ones by considering the entire pressure contributions to be subsumed within the first pressure gradient term. Resolving the combined pressure contribution in this alternative route would, however, require the consideration of the transverse-direction momentum equation akin to semi-analytical treatments of diffusioosmotic phenomena (see, for example, Ref. \cite{2006JCISKeh, 2007LangmuirKeh}) - a step which is by-passed in the present formalism. Next, the general expression for the contribution of the Maxwell's Stress to the body force is (assuming constant permittivity):
\begin{eqnarray}
\mathbf{F}_E = \epsilon \nabla^2\phi \nabla \phi,
\end{eqnarray}
where the total potential, $\phi$, is the sum of potentials due to streaming potential, $\phi_0$, and the potential screening the surface charge, $\psi$, so that $\phi = \phi_0 + \psi$. Further, noting that $\nabla^2 \phi \approx \partial_y^2 \psi$ because $\phi_0$ is constant along the transverse direction and both $\phi_0$ and $\psi$ are assumed to weakly vary along the x-direction, we have $\mathbf{F}_E = \epsilon \partial_y^2 \psi \nabla (\phi_0 + \psi)$. Additionally, since $\nabla \phi_0 = -\mathbf{E}_s$, we can write for the axial component of this force as:
\begin{eqnarray}
{F_{E_x}} = \epsilon \left( \frac{\partial ^2 \psi}{\partial y^2}\right) \left(-E_s + \frac{\partial \psi}{\partial x}\right), \label{eq:FE_ax}
\end{eqnarray}
%
%
%
Using Eqs. \ref{eq:FO_ax} and \ref{eq:FE_ax}, the x-momentum equation may then be written as:
\begin{widetext}
\begin{eqnarray}
0 = \frac{{ - dP}}{{dx}} + \mu \frac{{{\partial ^2}u}}{{\partial {y^2}}} + \mu \frac{{{\partial ^2}u}}{{\partial {x^2}}} - {k_B}({n_ + } + {n_ - })\frac{{\partial T}}{{\partial x}} - {k_B}T\frac{{\partial ({n_ + } + {n_ - })}}{{\partial x}} - \varepsilon \frac{{{\partial ^2}\psi }}{{\partial {y^2}}}{E_s} + \varepsilon \frac{{{\partial ^2}\psi }}{{\partial {y^2}}}\frac{{\partial \psi }}{{\partial x}} \label{eq:dimNS}
\end{eqnarray}
\end{widetext}

\subsection{Energy Equation}
Taking into consideration the effects of axial conduction and viscous dissipation the thermal transport equation can be written as:
\begin{eqnarray}
\rho C_p u \frac{\partial T}{\partial x} = k\frac{\partial^2T}{\partial y^2} + k\frac{\partial^2 T}{\partial x^2} + \mu \left(\frac{\partial u}{\partial y}\right)^2  \label{eq:Energy}
\end{eqnarray}
where $\rho$ is the density, $C_p$ is the specific heat capacity, $k$ is the thermal conductivity and $\mu$ is the viscosity of the electrolyte. In an effort to simplify the analysis without sacrificing the essential physics, we consider the thermo-physical properties to be temperature-invariant. This consideration also helps us to isolate the demonstrated effects from electrothermal effects that may occur as a consequence of variation of electrical properties with temperature.

\section{Dimensionless Forms}
To simplify numerical implementation, the governing equations are rewritten in dimensionless form using characteristic parameters of the system: the channel half-height $H$, the length of the channel $L$, the ionic concentration $n_0$  of the bulk electrolyte, the reference temperature $T_0$, and the characteristic velocity $u_{ref}$. Additionally, the potential is non-dimensionalized by the thermal voltage scaled by a factor of 4, $k_BT_0/4ez$ identical to Ref. \cite{2008PREChakrabortyDas}, and the cationic and the anionic diffusion constants are assumed equal so that $D_+ = D_- = D$. The aspect ratio of the channel geometry is denoted by $\alpha = H/L$. The new dimensionless variables (denoted by a tilde on top) are given by:
\begin{eqnarray}
&&\tilde{\psi} = \frac{ez\psi}{4k_BT_0}, \quad \tilde{u}=\frac{u}{u_{ref}}, \quad \tilde{n}_\pm = \frac{n_\pm}{n_0}, \nonumber \\ 
&&\quad \tilde{T}=\frac{T}{T_0}, \quad \tilde{x}=\frac{x}{L}, \quad \tilde{y}=\frac{y}{H}
\end{eqnarray}
\subsection{Dimensionless Poisson Equation} 
Using the aforementioned non-dimensionalization scheme in Eq. \ref{eq:Poisson}, we get the dimensionless Poisson equation as:
\begin{eqnarray}
\alpha^2 \frac{\partial^2 \tilde{\psi}}{\partial \tilde{x}^2} + \frac{\partial^2 \tilde{\psi}}{\partial \tilde{y}^2} = - \frac{1}{8} K^2 \left( \tilde{n}_+ - \tilde{n}_- \right), \label{eq:nondimPB}
\end{eqnarray}
where $K = H/\lambda$ and $\lambda = \sqrt{(\epsilon k_B T_0)/(2 n_0 e^2 z^2)}$ is the Debye length representing the characteristic thickness of the EDL so that $K$ represents the penetration of the EDL into the bulk relative to the channel half-height.

\subsection{Dimensionless Species Transport Equations}
Again using the aforementioned non-dimensionalization scheme in Eqs. \ref{eq:nplus_tp} and \ref{eq:nminus_tp}, the dimensionless transport equations for positive and negative ions become:
\begin{widetext}
\begin{eqnarray}
0 &=&  - \left( {\alpha^2 \frac{{{\partial ^2}{{\tilde n }_ + }}}{{\partial {{\tilde x }^2}}} + \frac{{{\partial ^2}{{\tilde n }_ + }}}{{\partial {{\tilde y }^2}}}} \right) + \frac{\partial}{\partial \tilde y}\left[ {\left( - \frac{4}{\tilde T}\frac{{\partial \tilde \psi  }}{{\partial \tilde y }} - \mathcal{Q} \frac{{\partial \tilde T }}{{\partial \tilde y }}\frac{1}{{{{\tilde T }^2}}} \right){{\tilde n }_ + }} \right] \nonumber      \\
&& \hspace{2cm} + \alpha^2 \frac{\partial }{{\partial \tilde x }}\left[ {\left( { - \frac{4}{{\tilde T }}\frac{{\partial \tilde \psi  }}{{\partial \tilde x }} + \tilde{E}_s - \mathcal{Q} \frac{{\partial \tilde T }}{{\partial \tilde x }}\frac{1}{{{{\tilde T }^2}}}} \right){{\tilde n }_ + }} \right] + \alpha Pe  \frac{\partial \left( \tilde{n}_+ \tilde{u} \right) }{\partial \tilde{x}}, \\
0 &=&  - \left( {\alpha^2 \frac{{{\partial ^2}{{\tilde n }_ - }}}{{\partial {{\tilde x }^2}}} + \frac{{{\partial ^2}{{\tilde n }_ - }}}{{\partial {{\tilde y }^2}}}} \right) + \frac{\partial }{\partial \tilde y}\left[ {\left( {\frac{4}{{\tilde T }}\frac{{\partial \tilde \psi  }}{{\partial \tilde y }} - \gamma \mathcal{Q}\frac{{\partial \tilde T }}{{\partial \tilde y }}\frac{1}{{{{\tilde T }^2}}}} \right){{\tilde n }_ - }} \right] \nonumber \\
&& \hspace{2cm} + \alpha^2 \frac{\partial }{{\partial \tilde x }}\left[ {\left( {\frac{4}{{\tilde T }}\frac{{\partial \tilde \psi  }}{{\partial \tilde x }} - \tilde{E}_s - \gamma \mathcal{Q} \frac{{\partial \tilde T }}{{\partial \tilde x }}\frac{1}{{{{\tilde T }^2}}}} \right){{\tilde n }_ - }} \right] + \alpha Pe \frac{\partial \left( \tilde{n}_- \tilde{u} \right) }{\partial \tilde{x}},
\end{eqnarray}
\end{widetext}
where $Pe = u_{ref}H/D$ is the Peclet number, $\mathcal{Q}=Q_+/k_B T_0$ and $\gamma = Q_-/Q_+$. The contribution to the advective terms from the transverse direction velocity has been assumed to be negligible. 

\subsection{Dimensionless Streaming Potential Field}
The dimensionless form of the streaming potential field using the aforementioned non-dimensionalization scheme in Eq. \ref{eq:Es} is:
\begin{eqnarray}
\tilde{E}_s = \frac{-\frac{1}{\alpha} Pe I_1 + I_2 + I_3 + \mathcal{Q} I_4}{I_5},
\end{eqnarray}
where the streaming potential field has been non-dimensionalized by $k_B T_0/ e z L$, and the expressions of the various integrals are:
\begin{eqnarray}
I_1 &=& \int_0^2 \left( \tilde{n}_+ - \tilde{n}_- \right) \tilde{u} d\tilde{y}, \\
I_2 &=& \int_0^2 \frac{\partial}{\partial \tilde{x}} \left( \tilde{n}_+ - \tilde{n}_- \right) d\tilde{y}, \\
I_3 &=& \int_0^2 4 \left( \frac{\tilde{n}_+ + \tilde{n}_-}{\tilde{T}} \right) \frac{\partial \tilde{\psi}}{\partial \tilde{x}} d\tilde{y}, \\
I_4 &=& \int_0^2 \left( \frac{\tilde{n}_+ - \gamma \tilde{n}_-}{\tilde{T}^2} \right) \frac{\partial \tilde{T}}{\partial \tilde{x}} d\tilde{y}, \\
I_5 &=& \int_0^2 \left( \frac{\tilde{n}_+ + \tilde{n}_-}{\tilde{T}} \right) d\tilde{y}.
\end{eqnarray}

\subsection{Dimensionless Navier-Stokes Equation}
The dimensionless Navier-Stokes equation using the aforementioned non-dimensionalization scheme in Eq. (\ref{eq:dimNS}) is:
\begin{eqnarray}
0 = 2 + \alpha^2 \frac{\partial^2 \tilde{u}}{\partial \tilde{x}^2} + \frac{\partial^2 \tilde{u}}{\partial \tilde{y}^2} - C \frac{\partial}{\partial \tilde{x}} \left( \tilde{n} \tilde{T} \right) \nonumber \\ \hspace{2cm} - \frac{8C}{K^2} \left( \tilde{E}_s - 4 \frac{\partial \tilde{\psi}}{\partial \tilde{x}}  \right) \frac{\partial^2 \tilde{\psi}}{\partial \tilde{y}^2},
\end{eqnarray}
where $C= \frac{2n_0 k_B T_0}{- L dP/dx}$ represents the strength of the osmotic pressure relative to that of the hydrodynamic pressure.

\subsection{Dimensionless Energy Equation}
Finally, the dimensionless energy equation by using the non-dimensionlization scheme in Eq. \ref{eq:Energy} is:
\begin{eqnarray}
\alpha Pe_T \tilde{u} \frac{\partial \tilde{T}}{\partial \tilde{x}} = \frac{\partial^2 \tilde{T}}{\partial \tilde{y}^2} + \alpha^2 \frac{\partial^2\tilde{T}}{\partial \tilde{x}^2} + Br_R \left(\frac{\partial \tilde{u}}{\partial \tilde{y}}\right)^2, \nonumber \\
\end{eqnarray}
where $Pe_T=\rho C_p u_{ref} H/k$ is the thermal P\'eclet number, and $Br_R = \mu u_{ref}^2/kT_0$ is a Brinkman number based on the reference temperature $T_0$.

\section{Boundary Conditions}
\subsection{Wall}
The channel wall is assumed to be at a constant zeta potential ($\psi=\zeta$) with no flux of ions across it  ($\hat{n} \cdot \mathbf{j}_\pm = 0$), with $\hat{n}$ depicting the unit vector normal to the surface. No-slip boundary condition is assumed: $u=0$. A linear temperature gradient is applied along the channel wall ($T=T_1+(T_f - T_1)x/L$). In dimensionless form these boundary conditions at the wall ($\tilde{y}=0$ and $0\le\tilde{x}\le1$) are expressed as:
\begin{eqnarray}
\tilde{\psi}&&=\tilde{\zeta}, \quad \frac{\partial \tilde{n}_\pm}{\partial \tilde{y}} \pm 4\left( \frac{n_\pm}{\tilde{T}} \frac{\partial \tilde{\psi}}{\partial \tilde{y}} \right) + \frac{Q_\pm}{k_B T_0} \left( \frac{\tilde{n}_\pm}{\tilde{T}^2} \frac{\partial \tilde{T}}{\partial \tilde{y}} \right) = 0, \nonumber \\
\tilde{u}&&=0, \quad \tilde{T} = \tilde{T}_1 \left\{ 1 + \tilde{x} \left( T_{ratio} - 1 \right) \right\},
\end{eqnarray}
where $T_{ratio} = T_f/T_1$.
\subsection{Centre-line}
At the channel centre-line we assume far-stream condition and set the electrokinetic potential to zero ($\psi=0$) in tandem with the electro-neutrality condition: $n_\pm=n_0$. Exploiting the symmetry of the channel about the centre-line, we set the gradient of the velocity and the temperature in the transverse direction to be zero: $\partial u/ \partial y$ and $\partial T/ \partial y = 0$. In non-dimensional form, these boundary conditions at the centre-line ($\tilde{y}=1$ and $0\le\tilde{x}\le1$) are given by:
\begin{eqnarray}
\tilde{\psi}=0, \quad \tilde{n}_\pm = 1, \nonumber \\
\frac{\partial \tilde{u}}{\partial \tilde{y}}=0, \quad \frac{\partial \tilde{T}}{\partial \tilde{y}} = 0.
\end{eqnarray}
\subsection{Entrance}
The fluid is assumed to enter the channel at the reference temperature ($T=T_0$) at any arbitrary velocity (say, $u=0.1 u_{ref}$). Assuming reservoir at the entrance we set ion concentration equal to the bulk concentration ($n_\pm=n_0$). We also assume that the induced potential is zero ($\psi=0$). In non dimensional form the boundary conditions at the entrance ($0\le\tilde{y}\le1$ and $\tilde{x}=0$) are given by:
\begin{eqnarray}
\tilde{\psi}=0, \quad \tilde{n}_\pm = 1, \quad \tilde{u}=0.1, \quad \tilde{T}=1.
\end{eqnarray}

\subsection{Exit}
When the fluid leaves the channel it is hydrodynamically fully developed. Again, assuming the presence of a reservoir at the exit, we set the ion concentration to the bulk value, $n_\pm=n_0$. We also assume that the induced potential is zero, $\psi=0$. The temperature profile at the exit is derived from energy conservation principle (i.e. energy leaving the control volume is the sum of energy entering the control volume and the energy generated through viscous dissipation); this conservation principle is expressed in dimensional form as:
\begin{widetext}
\begin{eqnarray}
\underbrace {\rho {C_p}\int\limits_0^H {u(0,y)T(0,y)dy}  - k{{\int\limits_0^L {\left. {\frac{{\partial T}}{{\partial y}}} \right|} }_{y = 0}}dx}_{{\text{Energy entering }}} + \underbrace {{\int\limits_0^H \int\limits_0^L \mu \left(\frac{{du}}{{dy}}\right)^2}dxdy}_{{\text{Energy generated}}} = \underbrace {\rho Cp\int\limits_0^H {u(L,y)T(L,y)dy} }_{{\text{Energy leaving}}}
\end{eqnarray}
\end{widetext}
In non-dimensional form, the boundary conditions at the exit ($0\le\tilde{y}\le1$ and $\tilde{x}=1$):
\begin{widetext}
\begin{eqnarray}
\tilde{\psi}=0, \quad \tilde{n}_\pm = 1, \quad \frac{\partial \tilde{u}}{\partial \tilde{x}} = 0, \nonumber \\
\alpha Pe_T \int\limits_0^1 {\tilde u \left( {0,\tilde y } \right)} \tilde T \left( {0,\tilde y } \right)d\tilde y  - {\int\limits_0^1 {\left. {\frac{{\partial \tilde T }}{{\partial \tilde y }}} \right|} _{\tilde y  = 0}}d\tilde x  + Br_R {\int\limits_0^1 \int\limits_0^1 \left( {\frac{{d\tilde u }}{{d\tilde y }}} \right)^2}d\tilde x d\tilde y  =  \alpha Pe_T \int\limits_0^1 {\tilde u \left( {1,\tilde y } \right)} \tilde T \left( {1,\tilde y } \right)d\tilde y. \label{eq:BC_exit_integral}
\end{eqnarray}
\end{widetext}
%

\vspace{5cm}
\section{Results and Discussions}
\begin{figure*}
\includegraphics[width=8cm]{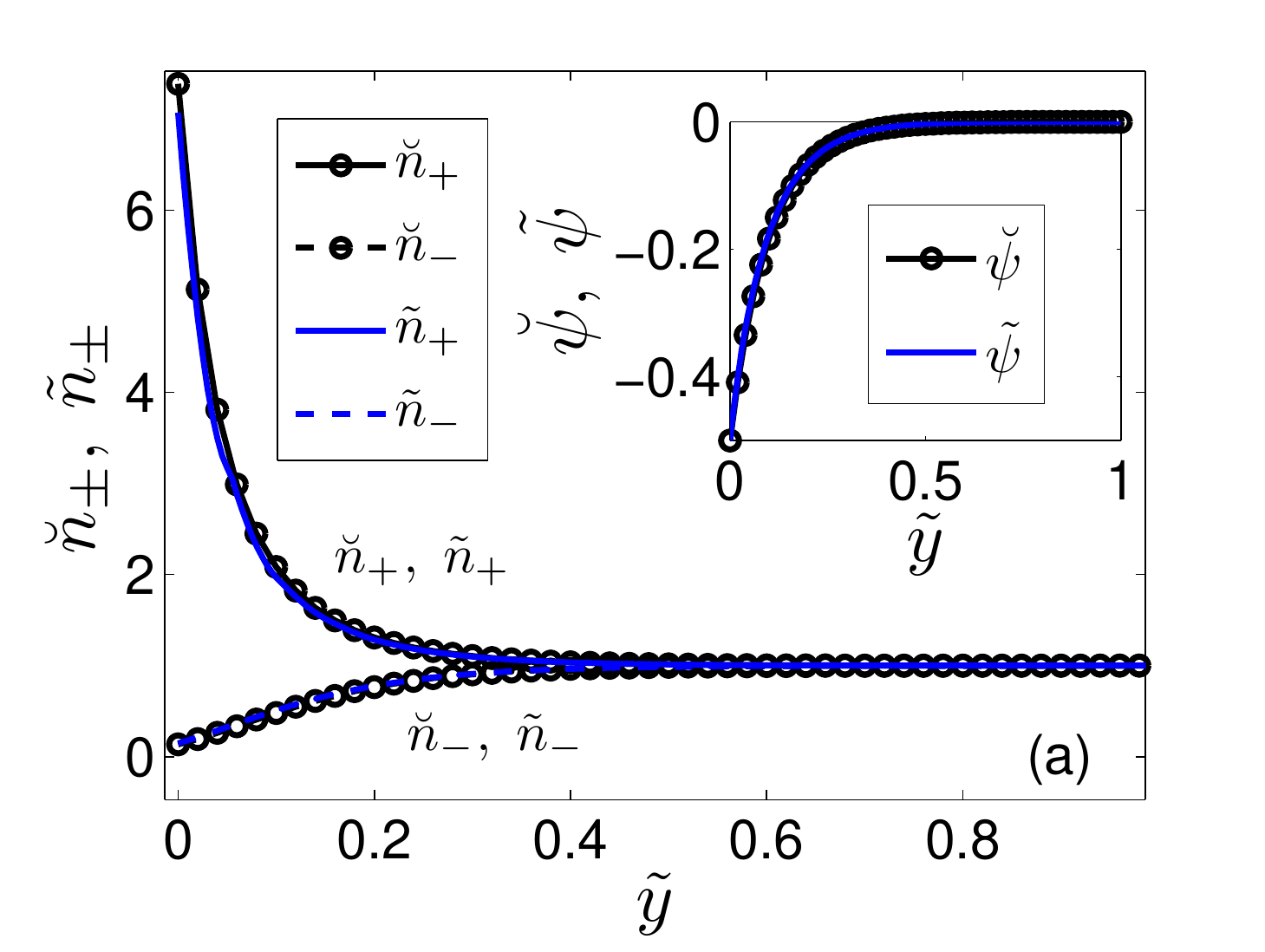}
%
\includegraphics[width=8cm]{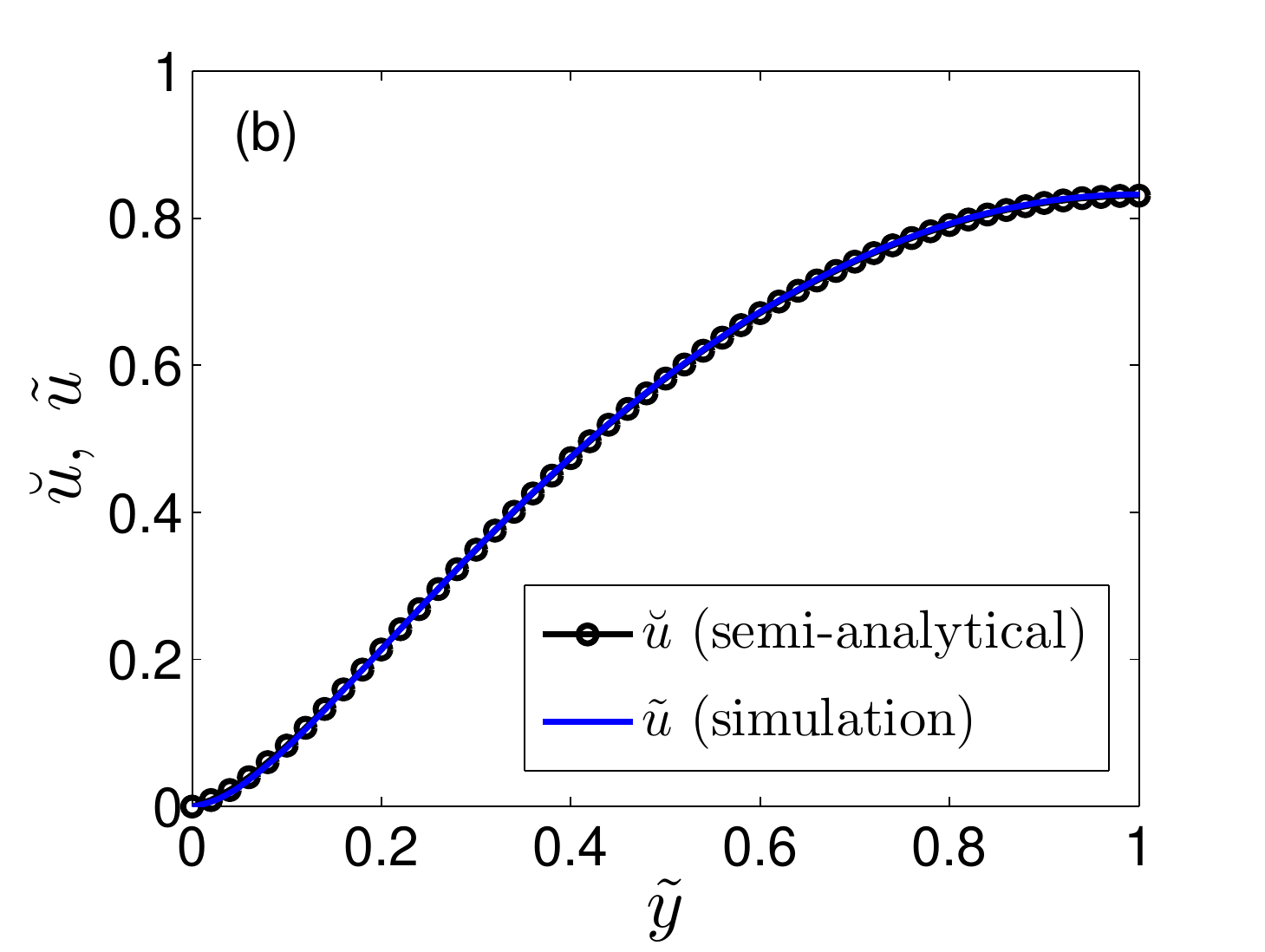}
\caption{(Left) Numer density of counterions and coions along the cross-section at $\tilde{x}=0.5$ corresponding to $\breve{\zeta}=-0.5$ and $K=10$ and in the absence of any temperature gradient. Inset shows the corresponding electrokinetic screening potential profile. (Right) The corresponding axial direction velocity profile. The plots with markers represent the results from the semi-analytical formulation while those without represent the results from the numerical framework developed in this study. These results are found to be in excellent agreement.}
\end{figure*}
The primary objective of the present work is to investigate the influence of temperature gradients on streaming potential mediated flows. The framework required to model such scenarios as represented through Eqs. \ref{eq:nondimPB}-\ref{eq:BC_exit_integral} is a weave of intrinsically coupled fields. Even with a number of simplifying assumptions this framework is manifestly more involved than those used in prior works which study streaming potential mediated flows in a number of different settings albeit, without exception, in isothermal conditions. Without further simplifications and/or approximations, this framework is not amenable to analytical treatments and, hence, requires the use of numerical techniques. For our purpose, we choose the finite element method as implemented in the COMSOL Multiphysics environment. Before setting out to report the results of the temperature gradient influences, we first establish the validity of our numerical framework by way of comparison of the isothermal condition results obtained from the same with those from a well-established semi-analytical formulation found in Ref. \cite{2008PREChakrabortyDas}; pertinently, this formulation has formed the basis of a number of subsequent works dealing with streaming potential mediated flows \cite{2009LangmuirSiddharthaDaSumanSir, 2009JChemPhysTamalDaSiddharthaDaSumanSir, 2009POFFarooqueSumanSir, 2010EPGaraiSumanSir, 2010LangmuirPrakashDaSumanSir, 2010POFJeevanSumanSir, 2011POFJeevanSumanSir, 2012EPJeevanSBRSirSumanSir, 2011LangmuirAdityaSumanSir, 2012APLAdityaSumanSir}. Next, we show the importance of considering the Soret effect by comparing results with those obtained from a consideration of temperature gradient influences solely through explicit manifestation of such gradients in the momentum equation. Finally, we explicate the specifics of these Soret effects through their influences on the ionic species flux on temperature gradient mediated streaming potential flows. Unless otherwise mentioned, we show all results corresponding to constant values of $\zeta = -0.5$, $K=10$, $C=0.828$, $\alpha=2.39 \times 10^{-6}$ and $Pe=6.85 \times 10^{-4}$.

\subsection{Validation of simulation results}

 In particular, for the sake of this validation study, we first switch off the influence of the temperature gradient in our numerical framework by setting $T_{ratio}=1$. Next, we restate the dimensionless axial direction velocity from the semi-analytical formulation of Ref. \cite{2008PREChakrabortyDas} in a suitably modified non-dimensional format to make the representation amenable for meaningful comparison as:
\begin{eqnarray}
\breve{u} = \left( 2\tilde{y} - \tilde{y}^2 \right) - 8\frac{C \breve{E}_s}{K^2} \tilde{\zeta} \left( 1 - \frac{\tilde{\psi}}{\tilde{\zeta}} \right).
\end{eqnarray}
Here the dimensionless streaming potential, $\breve{E}_s$ is given by:
\begin{eqnarray}
\breve{E}_s = \frac{\frac{1}{\alpha} Pe I_{1s}}{I_{2s} - 8\frac{C Pe}{\alpha K^2}I_{3s}},
\end{eqnarray}
where the expressions of the three integrals are:
\begin{eqnarray}
I_{1s} &=& \int_0^2 \left(\breve{n}_+ - \breve{n}_-\right) \left( \tilde{y}^2 - 2\tilde{y} \right) \; d\tilde{y}, \\
I_{2s} &=& \int_0^2 \left(\breve{n}_+ + \breve{n}_-\right) \; d\tilde{y}, \\
I_{3s} &=& \int_0^2 \breve{\zeta} \left(\breve{n}_+ - \breve{n}_-\right) \left( 1 - \frac{\breve{\psi}}{\breve{\zeta}} \right) \; d\tilde{y}.
\end{eqnarray}
It is to be noted that in this formulation, the electrokinetic potential profile is obtained by solving the Poisson equation:
\begin{eqnarray}
\frac{\partial^2 \breve{\psi}}{\partial \tilde{y}^2} = -\frac{1}{8} K^2 \left( \breve{n}_+ - \breve{n}_- \right),
\end{eqnarray}
where the ionic number densities are assumed to follow the Boltzmann distribution: $\breve{n}_\pm = \exp(\mp 4\breve{\psi})$.

In Fig. 2 (left panel), we show the distribution of the number density of the counterions and the coions along the cross-section of the channel at $\tilde{x}=0.5$ obtained from both the numerical implementation and the semi-analytical formulation. We also show, in the inset of Fig. 2, the comparison of the electrokinetic potential profile, accompanying the ion distribution, obtained from the two methods. Furthermore, in Fig. 2 (right panel), we show the velocity profiles corresponding to the distributions shown in the left panel again obtained from the two methods. All the results from the two methods are found to be in excellent agreement. This, therefore, sets a robust ground for using this numerical framework in further investigations of the influence of the temperature gradient. 

\subsection{Influence of temperature gradient}

\begin{figure}[h]
\includegraphics[width=8cm]{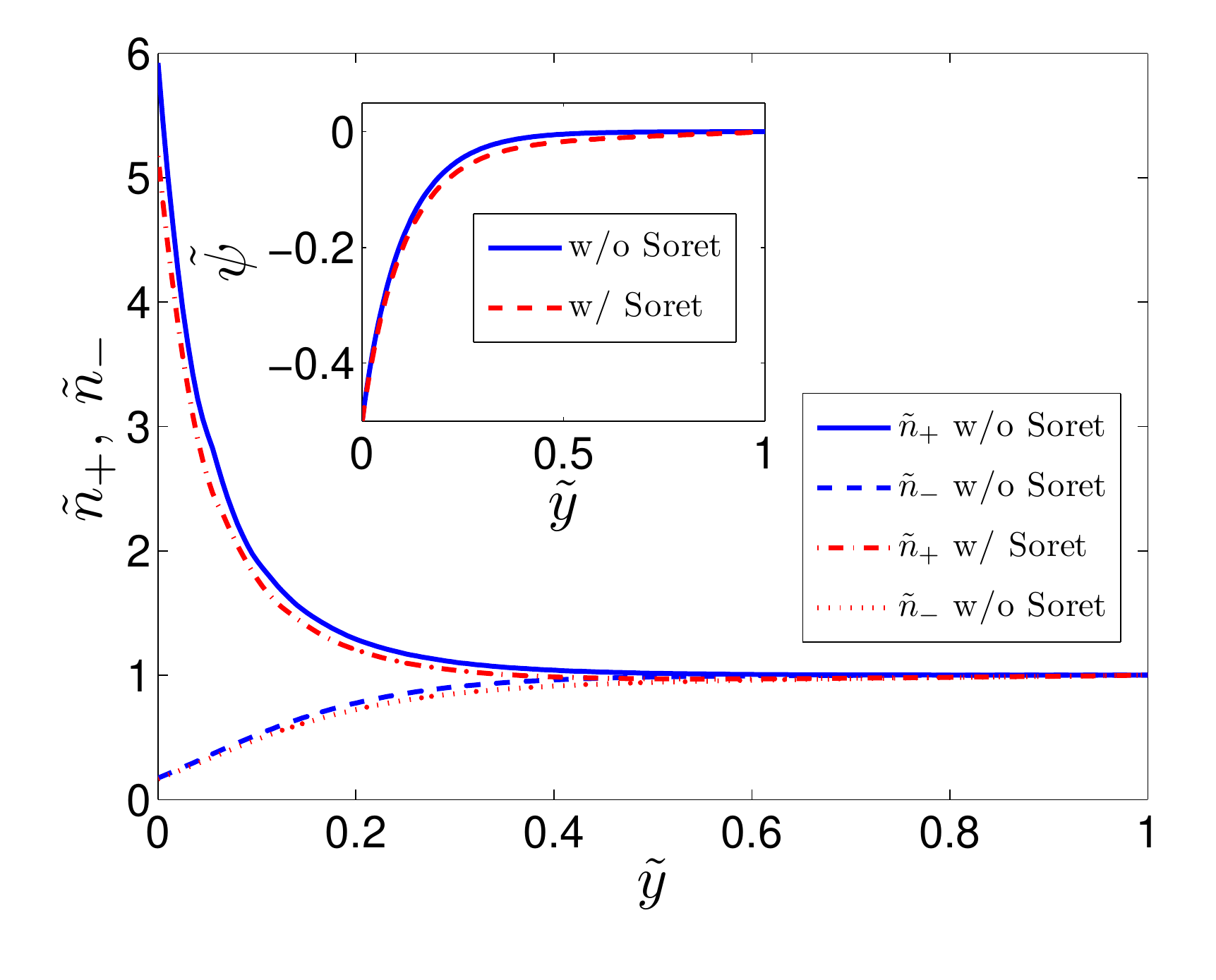}
\caption{Comparison of the numer density of counterions and coions considering Soret effect and that without along the cross-section of the channel at $\tilde{x}=0.5$ with $\tilde{\zeta}=-0.5$, $K=10$, $\mathcal{Q}=1.388$ and $\gamma=0.153$. Inset shows the corresponding comparison for electrokinetic screening potential profiles.} \label{fig:Fig3}
\end{figure}

\begin{figure}[h]
\includegraphics[width=8cm]{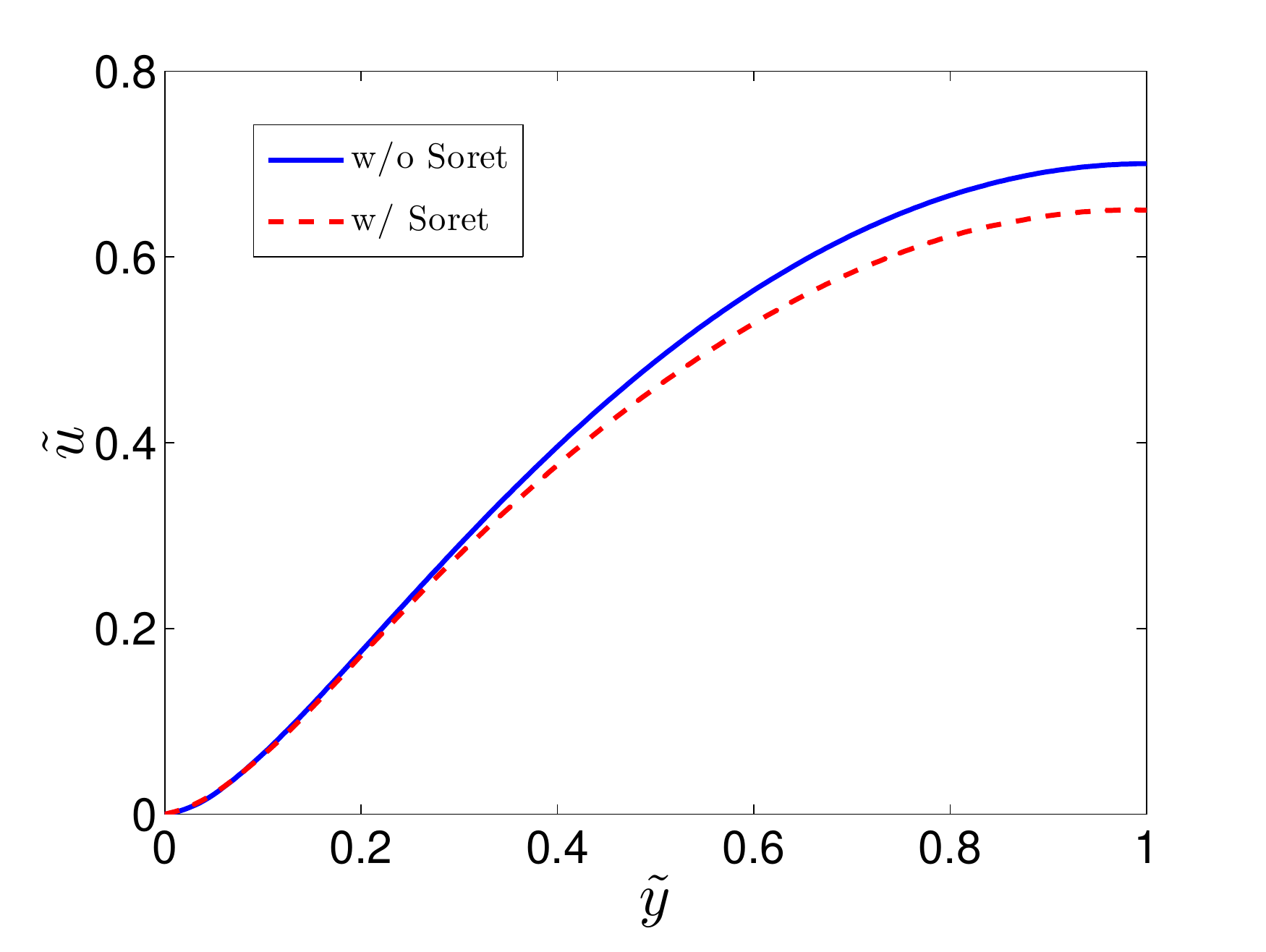}
\caption{Comparison of the velocity profiles considering Soret effect and that without along the cross-section of the channel at $\tilde{x}=0.5$ with $\tilde{\zeta}=-0.5$, $K=10$. Values of $\mathcal{Q}=1.388$ and $\gamma=0.153$ corresponding to a typical alkali halide \cite{2008PRLWurger} are used. These values will be subsequently referred to as $\mathcal{Q}^*$ and $\gamma^*$.} \label{fig:Fig4}
\end{figure}

The straightforward stratagem that one might expect to follow while investigating any influence of the temperature would be through the incorporation of such thermal gradients emanating from osmotic pressure contributions (which otherwise remain latent in conventional isothermal treatments) in the momentum equation. While such an expectation is not wrong it does not constitute the entire picture. For, even though the incorporation of temperature gradients in the momentum equation and the consequent influences on the velocity field may in turn be expected to influence the ionic flux leading to significantly coupled manifestation of such gradients in the overall field distributions, it still does not take into account the intrinsic dependence of the ionic species flux on the temperature gradient. Indeed, the influence of the temperature gradient on the ionic flux is very deeply ingrained having its basis on fundamental cross-couplings associated with the non-equilibrium thermodynamics of general species transport; in particular, the Soret effect. This determines the `diffusional' transport of a particle in response to an applied temperature gradient. 

As a first step in our discussion of the temperature gradient influence on the overall transport problem, we show, through Figs. 3 and 4, the differences in the results obtained from the consideration of Soret effect as compared to the aforementioned intuitive (but physically incomplete) stratagem. Fig. 3 shows the distribution of the number density of the counterions and the coions across the cross-section at $\tilde{x}=0.5$ while the inset shows the distribution of the potential profile across the same cross-section. Fig. 4 shows, again for the same cross-section, the velocity profile associated with the distributions in Fig. 3. The plots with Soret effects incorporated are obtained for $\mathcal{Q}=1.388$ and $\gamma = 0.153$. These values correspond to values of the ionic heat of transport of the cation, $Q_+$, and that of the anion, $Q_-$, typical for an alkali halide \cite{2008PRLWurger}; notably $Q_+ \sim 10 Q_-$. (In the ensuing discussion these values of $\mathcal{Q}$ and $\gamma$ will be referred to as $\mathcal{Q}^*$ and $\gamma^*$). While the influence of the Soret effects may not be evident in Fig. 3, it perceptibly clear in Fig. 4 where it is seen to suppress the magnitude of the velocity profile. This is particularly significant because it is the velocity profile which ultimately determines the volumetric flow rate and thus the throughput ratings of any micro- or nano-fluidic device. It is important to note that this suppression of the volumetric flow rate due to the temperature gradient influences is over and above that due to the streaming potential effects which invariably reduces the throughput in pressure-gradient driven flows through the generation of a back potential that drives a self-induced back electroosmotic flow. A discussion of the reason behind this suppression is in order. 

Positive values of $\mathcal{Q}$ and $\gamma$, originating from positive values of $Q_+$ and $Q_-$, represent thermophobic nature of the ions. This means that within the sole purview of the Soret effect (independent from any other electrokinetic or convective influences), these ions have a tendency to move from the hot to the cold region. Since temperature increases in the direction of the pressure-gradient driven flow (as shown in Fig. 1), the thermophoretic movement (associated purely with the Soret effect) of the ions is in the opposite direction. In perfect anlaogy with the physical explanation of the phenomenon of electroosmosis where it is the electrophoretic motion of the electrical double layer charges which get translated into a motion of the fluid, so also in the current situation, the back thermophoretic motion of the ions results, in turn, in a thermoosmotic flow of the fluid. Since this temperature-gradient mediated back flow of the fluid is in the same direction as the back electroosmosis-like flow associated purely with the induced streaming potential, the applied temperature gradient is seen to suppress the fluid flow to a greater extent compared to the case with no consideration of the Soret effects and, consequently, no possibility of any such intrinsic thermophoretic motion. There is, however, one effect which may countervail the suppression of the volumetric flow rate due to the Soret effect, and that is the thermoelectric effect. The genesis of this thermoelectric effect is, essentially, the same as that of the streaming potential as it arises due to the accumulation of the ions which get transported in response to the temperature gradient. In what follows, we explicate the combined interplay of the Soret effect and the thermoelectric effect through their influences on the volumetric flow rate.

\begin{figure}[h]
\includegraphics[width=8cm]{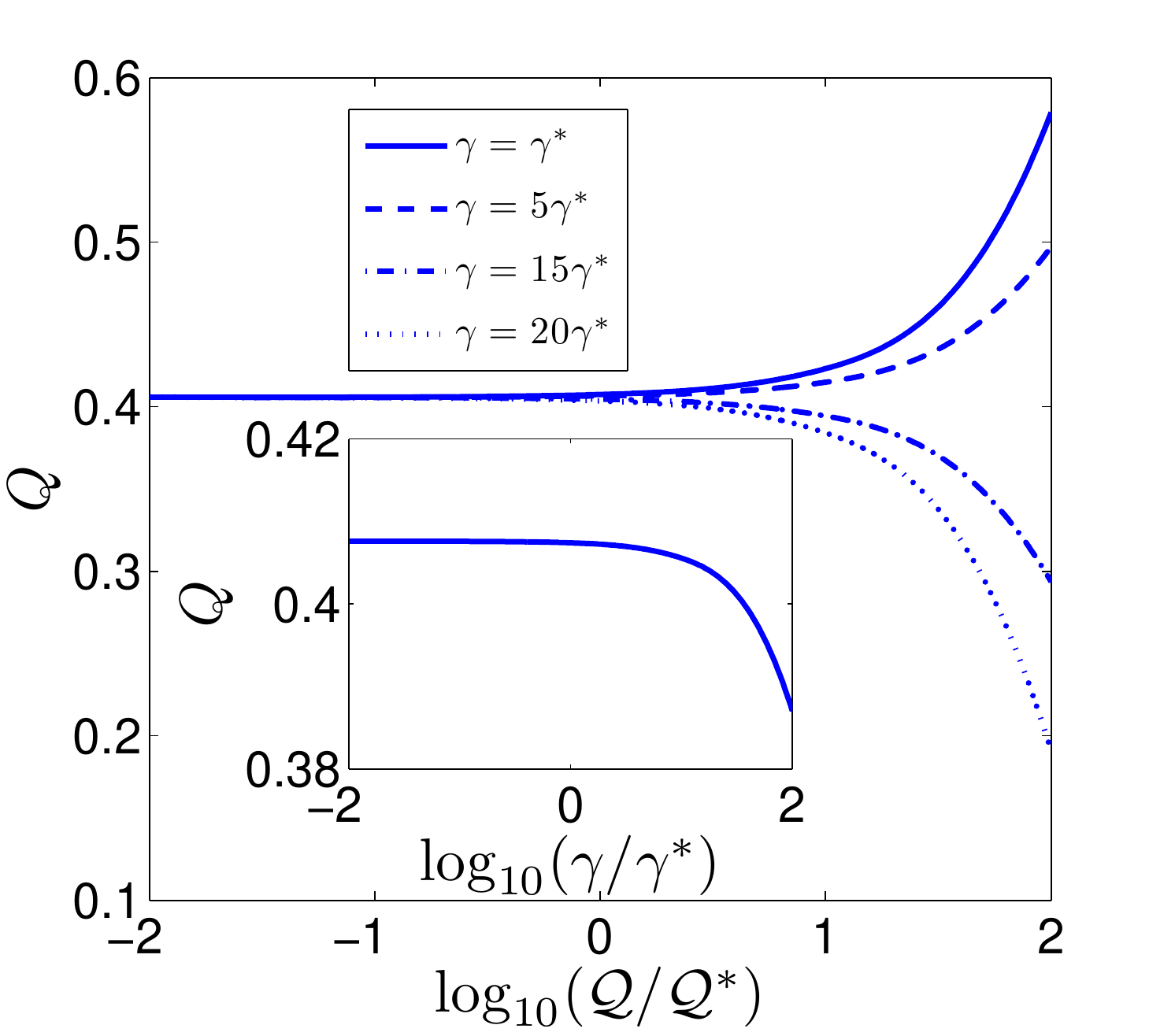}
\caption{Variation of the dimensionless volumetric flow rate corresponding to the variation of $\mathcal{Q}/\mathcal{Q}^*$ over four orders of magnitude for four different values of $\gamma/\gamma^*$. Inset shows the variation of the dimensionless volumetric flow rate with $\gamma/\gamma^*$ varying over four orders of magnitude.  The values of $\tilde{\zeta}=-0.5$, $K=10$ and $T_{ratio}=1.17$ are kept constant.} \label{fig:Fig5}
\end{figure}


In order to obtain a clearer picture of the combined influences of the electrokinetic, Soret and the as-yet unexplained thermoelectric effects, we investigate the extent to which the values of the Soret effect parameters $\mathcal{Q}$ and $\gamma$ affect the volumetric flow rate without limiting ourselves to the constraints of any specific electrolyte. In Fig. 5, we study the variation of the dimensionless volumetric flow rate over the half-channel cross-section (defined to be $\int_0^1 \tilde{u} d\tilde{y}$) as the value of $\mathcal{Q}$ is varied around $\mathcal{Q}^*$ over four orders of magnitude, i.e. $-2 \le \log_{10}(\mathcal{Q}/\mathcal{Q}^*) \le 2$ for different values of $\gamma$. Based solely on the explanation of the back flow of the ions (with the concomitant back flow of the fluid) presented earlier, it would be natural to expect that the volumetric flow rate would decrease with increase in the value of $\mathcal{Q}$. Contrary to that expectation, however, we observe that the value of the volumetric flow rate $Q$ is practically constant (the numerical value actually increases slowly) until about $\log_{10}\mathcal{Q}/\mathcal{Q}^* \sim 0.5$ beyond which it increases fast. This increase in the volumetric flow rate means that the suppressing influence of the Soret effect due only to the thermophobic back flow as discussed previously gets reduced. This is strongly indicative of the emergence of another physical factor. To motivate the understanding thereof, we note that in the case of the streaming potential the convective transport of the ions (in response to a pressure-gradient driven fluid flow) results in an accumulation of the ions downstream leading, in turn, to the generation of a back potential which suppresses the further transport of the ions and the fluid, thus negating the very cause which establishes it. In just the same way, the transport of the ions in response to the temperature gradient also leads to an accumulation of these ions albeit in the upstream direction. Again, just as in the case of the streaming potential, since the number density of the counterions is higher than that of the coions, there is a predominant accumulation of the counterions; this then leads to the generation of a thermoelectric field where the potential drops down along the direction opposite to that of the temperature gradient induced motion of the ions. In our case, this is also the direction along which the primary pressure gradient driven flow takes place. This thermoelectric field thus opposes the streaming potential field. Consequently, it also opposes the further flow of counterions in the upstream directin in response to both the Soret effect mediated thermophoresis and the streaming potential mediated back electroosmosis. This ultimately leads to a decrease in the suppression of the volumetric flow rate. 

A subtle point to note here is that besides the different number densities of the counterions and the coions, there is another factor which lies behind the genesis of the thermoelectric field. This factor is the unequal values of the ionic heat of transport of the counterions and the coions which essentially determines the extent of their thermophobicity. A lower value of the ionic heat of transport of the coions than that of the counterions (in our case the coions are the anions and the counterions are the cations), represented by $\gamma < 1$, ensures that the counterions have a greater propensity of getting accumulated in the upstream than the coions. It is only for the value $\gamma = 1$ (this is true for $\gamma \sim 10 \gamma^*$) that the thermoelectric field will be solely determined by the predominance in the number density of the counterions. The immediate deduction that we can make from this discussion of the influence of the relative values of the ionic heat of transport is this: as the difference in the values of the ionic heat of transport of the ions decreases, there will be a smaller manifestation of the thermoelectric effect. This is indeed seen to be so in Fig. 5 where the volumetric flow rate decreases for $\gamma=5\gamma^*$ implying that the thermoelectric field is now relatively weaker so that it cannot counteract the Soret effect mediated volumetric flow resistance. Further increase in the value of $\gamma$ beyond $10\gamma^*$ leads to a reversal in the nature of the thermoelectric field. Indeed, with $\gamma$ now being greater 1, it is the coions which have a greater propensity to flow upstream compared to the counterions: this leads to an inversion in the polarity of the thermoelectric field. Consequently, this thermoelectric field now aids the streaming potential mediated back electroosmotic flow leading to a significant decrease in the overall volumetric flow as seen in Fig. 5 for the plots corresponding to $\gamma=15\gamma^*$ and $\gamma=20\gamma^*$. Moreover, stronger the thermophobicity of the ions indicated by higher (positive) values of the ionic heats of transport, stronger is the Soret effect induced back flow and higher is the magnitude of the thermoelectric field: this explains the strong discrepancies in the values of the overall volumetric flow rate for higher values of $\log_{10}\mathcal{Q}/\mathcal{Q^*}$. The inset of Fig. 5 clearly shows the decreasing trend in the volumetric flow rate with increasing value of $\gamma$ relative to $\gamma^*$ for a constant value of $\mathcal{Q}=\mathcal{Q^*}$. An extremely point to remember is that no matter what the values of $\mathcal{Q}$ and $\gamma$ are, the combined consequences of the thermoelectric field, the Soret effect and the streaming potential mediated flows are fundamentally determined by the criterion that the net ionic current across any cross-section of the channel should necessarily be zero in the stationary state.    

\section{Summary and Conclusions}

In this study, we have presented the as-yet unaddressed consequences of temperature gradient on a streaming potential mediated pressure-gradient driven flow. We have incorporated the explicit dependence of this temperature gradient on the flux of the ions by considering the Soret effect in our modeling framework. To highlight the consequences of such consideration, we have first shown the differences between the cases with and without Soret effects. It is revealed that inclusion of Soret effects does indeed influence the velocity profiles quantitatively. Additionally, we have also studied the influence of the Seebeck effect which arises due to the differences in the ionic heats of transport of the cations and the anions. As the most important finding of our study, it is revealed that depending on the strength of the Seebeck effect and the concomitant thermoelectric field it generates due to separation of charges, the suppression of the volumetric flow rate due to the streaming potential field may be aided or opposed. Importantly, such an effect is manifested for an unchanged externally applied temperature gradient, and is, thus, a sole manifestation of the nature of the electrolyte.
 
These results have far-reaching practical consequences. Based on our findings, one may devise strategies in which temperature gradients may be employed for tuning the flow by aiding or opposing the streaming potential, depending on the relative dominance of Soret effect and thermoelectric effect. Since our studies are based on temperature-independent thermo-physical properties, the distinction between the new effects brought out in this work and electrothermal effects routinely addressed in microfluidics literature are clear. Moving further forward, one may exploit such flow control mechanisms for designing separation platforms based on a competing environment of thermodiffusion and electrophoresis. As delineated, these individual effects are not linearly super-imposable, as attributable to an intricate coupling between the thermal, potential, solutal, and velocity fields. The concerned implications, being significantly non-intuitive in nature, may hold the promises of addressing a new paradigm of microfluidic devices that relies on a strong thermo-electrical coupling with the aid of an imposed temperature gradient, without necessarily appealing to electrothermal principles. Our work further demonstrates that this subtle conjecture may be verified in complete absence of any externally applied electrical field, by appealing to the establishment of streaming potential fields in the presence of a simple pressure-driven flow.

\bibliography{sptemp}
\end{document}